\documentclass[nofootinbib,letter]{revtex4}

\usepackage{epsfig}
\newcommand{\beq}{\begin{equation}}
\newcommand{\eeq}{\end{equation}}
\newcommand{\beqa}{\begin{eqnarray}}
\newcommand{\eeqa}{\end{eqnarray}}
\newcommand{\bra}[1]{\ensuremath{{\langle{#1}|}}}
\newcommand{\ket}[1]{\ensuremath{{|{#1}\rangle}}}
\newcommand{\braket}[2]{\ensuremath{{\langle{#1}|{#2}\rangle}}}

\newcommand{\kket}[1]{\ensuremath{{|{#1}\rangle\!\rangle}}}
\newcommand{\bbrakket}[2]{\ensuremath{{\langle\!\langle{#1}|{#2}\rangle\!\rangle}}}
\newcommand{\average}[1]{\ensuremath{{\langle{#1}\rangle}}}

\begin{document}

\title{Loading a Bose-Einstein Condensate onto an Optical Lattice:\\ an Application of Optimal Control Theory to The Non Linear Schr\"odinger Equation} 
\author{Shlomo E. Sklarz}
\author{David J. Tannor}
\affiliation{Department of Chemical Physics, Weizmann Institute of Science \\
76100, Rehovot, Israel,\\ Tel 972-8-9343723, Fax 972-8-9344123}
\date{\today}
\begin{abstract}
Using a set of general methods developed by Krotov [A.~I. Konnov and V.~A. Krotov, Automation and Remote Control, {\bf 60}, 1427 (1999)], 
we extend the capabilities of Optimal Control Theory to the Nonlinear Schr\"odinger Equation (NLSE).  The paper 
begins with a general review of the Krotov approach to optimization.  Although the linearized version of the method
is sufficient for the linear Schr\"odinger equation, the full flexibility of the general method is required for 
treatment of the nonlinear Schr\"odinger equation.  Formal equations for the optimization of the
NLSE, as well as a concrete algorithm are presented.  As an illustration, 
we consider a Bose-Einstein condensate initially at rest in a 
harmonic trap.  A phase develops across the BEC when an optical lattice potential is turned on. The goal is to counter 
this effect and keep the phase flat by adjusting the trap strength. The problem is formulated in the language of
Optimal Control Theory (OCT) and solved using the above methodology.  To our knowledge, this is the
first rigorous application of OCT to the Nonlinear Schr\"odinger equation, a capability that is bound to have
numerous other applications.  
\end{abstract}
\maketitle
\section{\label{sec:Intro}Introduction}
In recent years much progress has been made in the use of Optimal Control Theory (OCT), to coherently control quantum mechanical systems governed by the Schrodinger equation. 
Such systems include controlled manipulations of molecular wave packets \cite{Rabitz, Gordon97, Rice2000}, high harmonic generation \cite{Murnane} and realization of quantum computing algorithms \cite{Rangan01}. 

In this paper for the first time the application of OCT is extended in a systematic way to systems governed by the NLSE, such as solitons in fiber optics and Bose-Einstien condesates (BEC's) in atomic physics. We begin with a general description of the Krotov iterative method \cite{Krotov84}. We describe first its application to quantum systems governed by the linear Schr\"odinger equation and then show how a generalized version of this method \cite{Konnov99} can be used to treat non-linear problems. 

Finally we consider a concrete problem governed by the NLSE, namely a BEC evolving under the Gross-Pitaevskii equation. The use of a BEC as a realization of quantum computing is widely being considered, as this is a macroscopic entity which nevertheless behaves quantum mechanically. The fact that a BEC carries  a definite phase which can be manipulated and controlled is a striking manifestation of this quality. For many computation applications it is desirable to split the BEC up into localized pieces each of which can then be viewed as a quantum bit and manipulated as such. This is achieved by the switching on of an optical lattice potential; however the switching on of the optical lattice causes a phase to accumulate across the BEC which is undesirable for use in computing applications. It is the cancellation of this effect which is the goal of this model problem.   

The outline of the paper is as follows: In section \ref{sec:KrotovMethod} the Krotov method is reviewed. Its application to linear problems in general and the Schr\"odinger equation in particular are discussed. Section \ref{sec:ApplNLSE} deals with the application of OCT to the NLSE problems and demonstrates this by solving the BEC problem mentioned above. Finally in section \ref{sec:Conclude} we conclude and suggest further applications.

\section{\label{sec:KrotovMethod}Krotov Method of Optimization}

\subsection{Description of problem}
Consider a state of some system which can be defined by a vector of variables $\psi$  and which is controlled by a set of variables $u$, through the state equations of motion 
\beq\label{eqKrtv1}
\dot\psi=f(t,\psi,u).
\eeq 
The initial value of  $\psi$, $\psi(0)=\psi_0$, is fixed but evolves over time to some final value $\psi(T)=\psi_T$. The history of evolving state vectors is called the state trajectory and the history of control input is termed the control history or just the control
\footnote{Here and throughout the following, vector treatment of the appropriate variables is assumed although all vector notation is omitted to avoid congestion. Multiplication of vector variables is therefore to be understood as a dot product. An indexed notation of vector components will be used only when unavoidable for clarity and summation convention will then be implied.}.

Given the trajectory $\psi(t)$ and the control $u(t)$, we define a 'process' $w=(\psi(t),u(t))$ as a pair of histories $\psi(t)$, $u(t)$ satisfying eq. (\ref{eqKrtv1}).  
We can now define the objective functional on the process $w$:
\beq
J[w]=F(\psi(T))+\int_0^T f^0(t,\psi(t),u(t))dt,
\eeq
where $F(\psi(T))$ and $f^0(t,\psi(t),u(t))$ are general functions that represent the dependence of $J$ on the terminal and intermediate time values of $\psi$ respectively.
It is required to find a process $w$ for which the objective  obtains its smallest value.

\subsection{Utility constructs and definitions}
For a continuously differentiable scalar function $\phi(t,\psi)$, we define the following functional
\beq
  L[w;\phi]=G(\psi_T)-\int^T_0R(t,\psi(t),u(t))dt-\phi(0,\psi_0),
\eeq
where
\beqa
  G(\psi_T)&=&F(\psi_T)+\phi(T,\psi_T),\\
  R(t,\psi,u)&=&\frac{\partial\phi}{\partial\psi}f(t,\psi,u)-f^0 (t,\psi,u)+{\partial\over\partial t}\phi(t,\psi).
\eeqa
The functions $R$ and $G$ are designed to  
separate out the dependence of $L[w;\phi]$ on the final time and intermediate time respectively.

It can be shown that for any function $\phi$ and process  $w=(\psi(t),u(t))$, $L[w;\phi]=J[w]$. The derivation goes as follows:
\begin{widetext}
\beqa
  L[w;\phi]&=&G(\psi_T)-\int^T_0R(t,\psi(t),u(t))dt-\phi(0,\psi_0)\nonumber\\
	&=&F(\psi_T)+\phi(T,\psi_T)-\int^T_0\left\{\frac{\partial\phi}{\partial\psi}f(t,\psi,u)-f^0 (t,\psi,u)+\frac{\partial\phi}{\partial t}\right\}dt-\phi(0,\psi_0)\nonumber\\
	&=&F(\psi_T)+\phi(T,\psi_T)-\int^T_0\left\{\frac{\partial\phi}{\partial\psi}\frac{\partial\psi}{\partial t}+\frac{\partial\phi}{\partial t}-f^0 (t,\psi,u)\right\}dt-\phi(0,\psi_0)\nonumber\\
	&=&F(\psi_T)+\phi(T,\psi_T)-\int^T_0\frac{d\phi}{dt}dt+\int^T_0f^0(t,\psi,u)dt-\phi(0,\psi_0)\nonumber\\
	&=&F(\psi_T)+\int^T_0f^0(t,\psi,u)dt=J[w]\label{eqKrtv10}.
\eeqa
\end{widetext}
Obviously therefore, minimizing $L[w;\phi]$  for any $\phi$ minimizes $J[w]$, and minimizing  $L[w;\phi]$ can be achieved by separately minimizing $G(\psi_T)$ and maximizing $R(t,\psi,u)$.

It is convenient for later reference to define the function $H$ through the following relation
\beq
R(t,\psi,u)\equiv H(t,\psi,u,\frac{\partial\phi(t,\psi)}{\partial\psi})+{\partial\over\partial t}\phi(t,\psi),
\eeq
where
\beq\label{eqKrtv6}
  H(t,\psi,u,p)=pf(t,\psi,u)-f^0(t,\psi,u).
\eeq
Note the extra parameter in $H$ denoted $p$, which emphasizes that $\psi$ and ${\partial\phi\over\partial\psi}$ should be treated as independent variables, with respect to $H$.
\subsection{An iterative algorithm}
We now return to our main goal and describe the Krotov iterative method for finding a sequence of processes $\{w_s\}$ which monotonically decrease the value of the objective $J[w]$ \cite{Krotov84, Krotov96}. The central idea is that as we have complete freedom in choosing the potential $\phi(\psi,t)$, we can construct $\phi$ such that our current estimate of the state history will {\it maximize} $L[w;\phi]$, and so become the worst of all possible histories. We are then free to find a new estimate for the control $u(t)$ which will minimize $L[w;\phi]$ with respect to its explicit dependence on $u(t)$, without worrying about the effect of $u(t)$ on $L[w;\phi]$ through the change of $\psi(t)$, as that can only be improved.

We begin by taking an arbitrary control history $u^0(t)$ and the corresponding state trajectory $\psi^0(t)$ which constitute together a process $w^0$.
\begin{enumerate}
  \item We first construct a function $\phi(t,\psi)$ such that $L[w;\phi]$ is a maximum with respect to $\psi(t)$ at the point $w^0$. This is equivalent to the following two conditions:	 
\beqa
  R(t,\psi^0(t),u^0(t))&=&\min_\psi R(t,\psi(t),u^0(t))\label{eqKrtv12}\\
  G(\psi^0_T)&=&\max_\psi   G(\psi_T),\label{eqKrtv13}
\eeqa
where the functions $R$ and $G$ are calculated using the new  $\phi(t,\psi)$. 
In other words we choose $\phi(t,\psi)$ such that our current $\psi^0(t)$ will be the worst of all possible $\psi(t)$'s in minimizing the objective $L[w;\phi]=J[w]$ (maximizing $R$, minimizing $G$). Any change in $\psi$ brought about by a new choice of $u(t)$ will now only improve the minimization of $J[w]$. (see figure \ref{Rphi}) 
\begin{figure}[htb]
\fbox{  \epsfig{figure=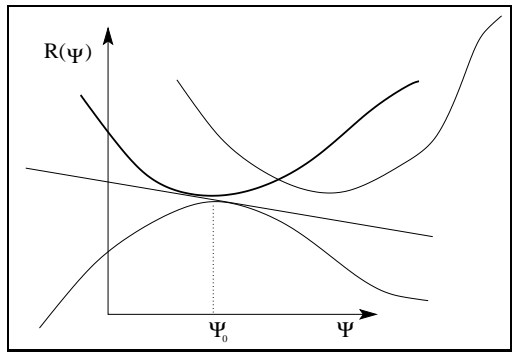,width=2.5in}}
   \caption[Rphi]{\label{Rphi} Sketch of various $R(\psi)$ determined by different choices of $\phi(t,\psi)$. The construction of $\phi(t,\psi)$ is the one that minimizes $R(\psi)$ at $\psi=\psi_0$, i.e. is the worst $R$ at the current $\psi_0$. }
\end{figure}
  \item For $\phi(t,\psi(t))$ we find a control $u(t)$ that maximizes $H(t,\psi,u,{\partial\phi \over \partial \psi})$ and denote it by
\beqa \label{eqKrtv11}
\tilde u(t,\psi)&=&arg \max_u H(t,\psi,u,{\partial\phi \over \partial \psi})\nonumber\\
                &=&arg \max_u R(t,\psi,u).
\eeqa
Note that the control $\tilde u(t,\psi)$ is still a function of $\psi$. This freedom will be removed in the next step.
  \item  We require that $\tilde u(t,\psi)$ and $\psi(t)$ be consistent with each other through the equations of motion. The equation of motion (\ref{eqKrtv1}) (with its initial conditions) together with the equation for the control $u=\tilde u(t,\psi)$ (\ref{eqKrtv11}), provide two equations for the two unknowns $u$ and $\psi$. These equations may be solved self consistently for $u$ and $\psi(t)$, obtaining the new process $w=(u,\psi)$.
  \item It is now guaranteed that minimization of the objective has been improved so that $J[w]<J[w^0]$; this completes the current iteration.
The new $w$ becomes a starting point for the next iteration, $w\to w^0$, and operations 1-3 can now be repeated to achieve further decrease in the objective.
\end{enumerate} 

We proceed to prove that indeed the new $J[w]\leq J[w^0]$.
First we assert using eq. (\ref{eqKrtv10}) that 
\beq
J[w^0]-J[w]=L[w^0;\phi]-L[w;\phi].
\eeq
Also 
\begin{widetext}
\beqa
L[w^0;\phi]-L[w;\phi] &=& G(\psi^0_T)-G(\psi_T)+\int_0^T\left\{R(t,\psi(t),u(t))-R(t,\psi^0(t),u^0(t))\right\}dt\nonumber\\
&=& \Delta_1+\Delta_2+\Delta_3,
\eeqa
\end{widetext}
where
\beqa
  \Delta_1 &=& G(\psi^0_T)-G(\psi_T),\label{eqKrtv13a}\\
  \Delta_2 &=& \int_0^T\left\{R(t,\psi(t),u(t))-R(t,\psi(t),u^0(t))\right\}dt,\label{eqKrtv14}\\
  \Delta_3 &=& \int_0^T\left\{R(t,\psi(t),u^0(t))-R(t,\psi^0(t),u^0(t))\right\}dt.\label{eqKrtv15}
\eeqa
The nonnegativeness of $\Delta_3$ and $\Delta_1$ follow from conditions (\ref{eqKrtv12}) and (\ref{eqKrtv13}) respectively, and by eq. (\ref{eqKrtv11}), the choice of a new control ensures the nonnegativeness of $\Delta_2$. This completes the proof.

\subsection{Construction of $\phi$ to 1st order in $\psi$}
In implementing the above iterative method the main difficulty lies in step 1. Here it is necessary to determine a function $\phi(t,\psi)$ that, by conditions (\ref{eqKrtv12}) and (\ref{eqKrtv13}), will ensure the absolute maximum and minimum of the functions $R$ and $G$ respectively on the trajectory $\psi_0(t)$, i.e. to choose $\phi(t,\psi)$ to give the worst possible $L[\psi_0;\phi]$. A necessary condition for an extremum of $R$ and $G$ at $w^0=(\psi^0,u^0)$ is the existence of a stationary point there, but in order to make the conditions  sufficient, it is necessary to add conditions of positivity and negativity  on the second derivatives of $R$ and $G$ respectively. 

We leave the additional requirements on the second derivatives for a later section and  restrict ourselves in this section solely to determining the conditions for a stationary point in  $R$ and $G$, which are as follows:
\footnote{In the derivation below the following delicate point should be noted: $R(t,\psi, u)$ is by definition a function of three variables, whereas $H(t,\psi, u, p)$ is a function of an additional argument $p=\partial\phi/\partial \psi$. In
other words although $p$ depends on $\psi$ with respect to $R$, this is not
the case with respect to $H$ where $\psi$ and $p$ are to be taken as independent variables. In eq. (\ref{eqKrtv16}) we wish to vary $R$ with respect to $\psi$ which means $H$ must be varied in both $\psi$ and $p$; $\partial R/\partial\psi\leftrightarrow\{(\partial p/\partial\psi)\partial/\partial p+\partial/\partial\psi\}H$.}
\begin{widetext}
\beqa
{\partial\over \partial \psi} R(t,\psi^0,u^0)&=&{\partial^2\phi(t,\psi^0)\over \partial \psi^2} f(t,\psi^0,u^0)+{\partial \phi\over \partial\psi}{\partial\over \partial \psi} f(t,\psi^0,u^0)-{\partial\over \partial \psi} f^0(t,\psi^0,u^0)+{\partial\over\partial t}{\partial \phi(t,\psi_0)\over \partial \psi}\nonumber\\*
&=&{\partial \over \partial\psi}H(t,\psi^0, u^0, \chi)+{\partial^2\phi(t,\psi^0)\over \partial \psi^2} f(t,\psi^0,u^0)+{\partial\over\partial t}{\partial \phi(t,\psi_0)\over \partial \psi}\nonumber\\*
&=&{\partial \over \partial\psi}H(t,\psi^0, u^0, \chi)+\left({\partial \psi\over \partial t}{\partial \over \partial\psi}+{\partial\over\partial t}\right){\partial \phi(t,\psi_0)\over \partial \psi}\nonumber\\*
&=&{\partial \over \partial\psi}H(t,\psi^0, u^0, \chi)+{d \chi\over dt}=0\label{eqKrtv16}\\
{\partial G(t,\psi_T^0,u^0)\over \partial \psi_T}&=&{\partial F(\psi_T^0,u^0)\over \partial \psi_T}+{\partial \phi(T,\psi_T^0)\over \partial\psi_T}\nonumber\\*
&=&{\partial F(\psi_T^0,u^0)\over \partial \psi_T}+\chi_T =0,
\label{eqKrtv17}
\eeqa
\end{widetext}
where\footnote
{We stress that in the following definition, $\chi(t)$ is a function solely of $t$ and is obtained by inserting the explicit dependence of $\psi^0(t)$ into $\partial \phi(t,\psi)/\partial\psi$. This explains the advance to the third line in  eq. (\ref{eqKrtv16}) where the total time derivative of  $\partial \phi(t,\psi)/\partial\psi|_{\psi^0}$ is translated to a simple time derivative of $\chi(t)$.}

\beq \label{eqKrtvChi}
\chi(t)={\partial \over \partial\psi}\phi(t,\psi^0(t)).
\eeq
Eq. (\ref{eqKrtv16}) and (\ref{eqKrtv17}) can be restated as an equation of motion for the vector $\chi$ and the definition of $H(t,\psi,u,p)$, eq. (\ref{eqKrtv6}), can be used to rewrite the equation of motion for $\psi$, eq. (\ref{eqKrtv1}), in the following  compact form
\beqa
\dot\chi &=& -{\partial \over \partial\psi}H(t,\psi^0,u^0,\chi)\nonumber\\
 &&\mbox{ with boundary conditions }\chi(T)=-{\partial F(\psi_T^0,u^0)\over \partial \psi_T}\label{eqKrtv19}\\
\dot\psi&=&{\partial \over \partial\chi}H(t,\psi,u^0,\chi)\nonumber\\
&&\mbox{ with boundary conditions }\psi(0)=\psi_0.\label{eqKrtv20}
\eeqa
These equations constitute a so called Hamiltonian system with a Hamiltonian $H(t,\psi,u,\chi)$, and the variables $\psi$ and $\chi$ are said to be conjugate.
Eq. (\ref{eqKrtvChi}) shows that the variable $\chi$ represents the function $\phi(t,\psi)$ to first order in $\psi$. Eqs. (\ref{eqKrtv19},\ref{eqKrtv20}) give the prescription for constructing $\chi(t)$. 

Creating the conjugate variable $\chi$ to fulfill the above requirements enforces a stationary point in $R$ and $G$ with respect to $\psi$. As explained above it is also necessary, in the general case, that the stationary point be an extremum and therefore that the second derivatives of $G$ and $R$ with respect to $\psi$ are negative and positive respectively. However  
for problems linear in $\psi$ it so happens that a stationary point is sufficient. This will be illustrated by a concrete example in the next section, after which we return to our main line of discussion completing the conditions for obtaining $\phi(t,\psi)$ in the general nonlinear case.

\subsection{A linear problem and application to the Linear Schr\"odinger equation optimization}
Consider a problem where the equations of motion are linear in the state variable
\beq\label{eqKrtv21}
\dot\psi=a(u)\psi, \quad F(\psi_T)=b\psi_T, \quad f^0=f^0(u).
\eeq
We proceed to show that it is sufficient to choose $\phi(t,\psi)=\chi(t)\psi$ to achieve monotonic increase in the objective at each iteration.
The Hamiltonian for this problem by the definition, eq (\ref{eqKrtv6}), is $H(t,\psi, u, \chi)=\chi a(u)\psi-f^0(u)$,
so we get by applying eq. (\ref{eqKrtv19}) 
\beq
\dot\chi=-a(u^0)\chi, \quad \mbox{ with boundary conditions } \chi(T)=-b.
\eeq
Using the above we find that 
\beqa
R(t,\psi,u^0)&=& \chi a(u^0)\psi-f^0+{\partial\chi \over \partial t}\psi=(\chi a(u^0)+{\partial\chi \over \partial t})\psi-f^0\nonumber\\
           &=&-f^0(u^0),\\
G(\psi_T)&=&b\psi_T+\chi_T\psi_T=(b+\chi_T)\psi_T=0,
\eeqa
which are independent of $\psi$. This implies that both $\Delta_1$ and $\Delta_3$ in eqs. (\ref{eqKrtv13a}) and (\ref{eqKrtv15}) vanish. By maximizing $\Delta_2$ (eq. (\ref{eqKrtv14})), the objective is guaranteed to decrease at each iteration. Note that in the linear case there is no need to check the second derivatives of $G$ and $R$ since, as $R$ is linear in $\psi$ and we set ${\partial R\over \partial\psi}|_{\psi^0}=0$, $R$ must be independent of $\psi$. (see figure \ref{Rphi}) Therefore the control $u$ can be made to maximize $R$ without the resulting change in $\psi(t)$ having any effect on the objective.

The above  example encompasses the problem of optimization of a quantum mechanical wave function  governed by the linear Schr\"odinger equation
\beq
\ket{\dot\psi}=-i\hat H(u)\ket{\psi}.
\eeq
Some care is necessary, however, if the objective takes the form $F(\psi_T)=-\bra{\psi_T}P\ket{\psi_T}$, which is not strictly linear in the state vector as in eq. (\ref{eqKrtv21}). Another complication arises from the fact that the state vector $\psi$ is an element of a complex Hilbert space. We therefore work this problem out in full, and show that nevertheless the above choice of $\phi$ is sufficient in these problems just as in the linear example, due to the fact that the target projection operator $P$ is positive definite. \cite{Krotov, Somloi93}

As above, we set $\phi(t,\psi)=\braket{\chi}{\psi}+\braket{\psi}{\chi}$ (where we have included the complex conjugate as an extra independent state variable) and thus get for the Hamiltonian of the problem
\beqa
H(t,\psi,u,\chi)&=&-i\braket{\chi}{\hat H|\psi}+i\braket{\psi}{\hat H^\dagger|\chi}-\lambda f^0(u)\nonumber\\
&=&2\Im \braket{\chi}{\hat H(u)|\psi}-\lambda f^0(u),
\eeqa
which using eq. (\ref{eqKrtv19}) yields for the conjugate vector
\beqa\label{eqKrtvChiL}
\ket{\dot\chi}&=&-i\hat H^\dagger(u^0)\ket{\chi},\nonumber\\
&&\mbox{with boundary conditions }\ket{\chi_T}=P\ket{\psi^0_T}.
\eeqa
Reinserting this equation and its complex conjugate into the formulas for $R$ and $G$  we have
\beqa
R(t,\psi,u^0)&=&-i\braket{\chi}{\hat H(u^0)|\psi}+i\braket{\psi}{\hat H^\dagger(u^0)|\chi}-\lambda f^0(u^0)\nonumber\\
&&+\braket{\dot\chi}{\psi}+\braket{\psi}{\dot\chi}\nonumber\\
&=&-\lambda f^0(u^0)\\
G(\psi_T)&=&-\braket{\psi_T}{P|\psi_T}+\braket{\chi_T}{\psi_T}+\braket{\psi_T}{\chi_T}\nonumber\\
&=&-\braket{\psi_T}{P|\psi_T}+\braket{\psi^0_T}{P|\psi_T}+\braket{\psi_T}{P|\psi^0_T}.
\eeqa
$R$ is independent of $\psi$ as above, which guarantees that $\Delta_3$ of eq. (\ref{eqKrtv15}) vanishes. $G$ is dependent on $\psi_T$; however  (denoting $\Delta\psi=\psi-\psi^0$) the  
positiveness of $\Delta_1=\braket{\Delta\psi_T}{P|\Delta\psi_T}$ (eq \ref{eqKrtv13a}) is always guaranteed due to the positiveness of $P$.
Alternatively, note that the second derivative of $G$, ${\partial^2 G\over\partial\psi_T\partial\psi_T^*}=-P<0$ is always negative due to the positiveness of $P$, assuring that the condition for a maximum of $G(\psi_0)$ is automatically met.
\footnote
{ Another way to map the quantum control problems onto the linear example is by formulating the Schr\"odinger equation in the density matrix formalism of Liouville space like so: $\kket{\dot\rho}=i\mathcal{L}_h\kket{\rho}$ with the objective $F(\psi_T)=\bbrakket{B}{\rho}$ where $B$ is some target state. In this form the equivalence to the linear case becomes self evident.
}

Another intricate point regarding these problems which is often missed, is the following. In many problems in quantum mechanics it so happens that the equations of motion are linear in the control variable $u$, namely the field. This means that strictly speaking there is no proper maximum in the Hamiltonian of the control system $H(t,\psi,u,\chi)$ with respect to $u$, and stage 2 of the algorithm (eq. (\ref{eqKrtv11})) cannot be properly fulfilled. This problem is often overcome by adding a penalty function, $\lambda f^0(u)$ quadratic in $u$ to the objective as implicitly indicated above. The physical interpretation of  this construction is that placing a penalty on the fluence of the field, constrains the algorithm to search out the optimal {\it direction} of $u$ rather than minimizing the objective by varying its magnitude. The price paid by this solution, however, is that the algorithm often exerts much effort into minimizing the superficial penalty part of the objective at the expense of the really required terminal part. 

An alternative way to overcome this problem is by noticing that the algorithm does not really require that the Hamiltonian be maximized by $u$ at each iteration. All that is really required is that the Hamiltonian be increased by the new choice of $u$, which is enough to ensure that $\Delta_2$ (eq. (\ref{eqKrtv13})) be nonnegative. The penalty function can therefore be dropped and at each iteration $u$ should be increased  $u\to u+\lambda^{-1}{\partial H\over \partial u}$, where $\lambda^{-1}$ is some {\it macroscopic} constant which can be chosen arbitrarily.\footnote
{ The choice of $\lambda^{-1}$ is not restricted by the algorithm, however some limit must exist through the physics and numerics of the problem. The $\lambda$ can also be adjusted at will during the optimization to improve convergence. The choice of the symbol $\lambda$ intentionally points to the similarity  of the role this parameter plays with that of the strength of the penalty imposed above and to the similarity of the resulting forms for choice of the new $u$. The difference ends up being solely that the proposed method takes $\lambda^{-1}{\partial H/\partial u}$ to be a correction to the old field instead of an entirely new choice.}  
This is not to be confused, despite the formal similarity, with the gradient methods where the correction to $u$ must always be small such that its effect on any change in $\psi$ will remain in the linear regime.
\subsection{Construction of $\phi$ to 2nd order in $\psi$}
As noted above, for an equation of motion non linear in the state variable, it is necessary to fulfill conditions on the second derivatives of $R$ and $G$ and therefore $\phi$ must be chosen to contain higher orders in $\psi$. We therefore take
\beq\label{eqKrtvPhiDef}
\phi(t,\psi)=\chi_i\psi_i+{1\over 2}\Delta\psi^*_i\sigma_{ij}(t)\Delta\psi_j
\eeq
where  $\Delta\psi_i=\psi_i-\psi^0_i$, and the functions $\sigma_{ij}$ are to be determined such as to obtain the required extrema in $R$ and $G$. 
The conditions supplementary to eq. (\ref{eqKrtv16}-\ref{eqKrtv17}), necessary for fulfilling eq. (\ref{eqKrtv12}-\ref{eqKrtv13}), are the following  system of differential inequalities: 
\beqa
d^2R&\geq&0,\quad d^2R=\Delta\psi^*_i{\partial^2R(t,\psi^0,u^0)\over \partial\psi^*_i\partial\psi_j}\Delta\psi_j,\\
d^2G&\leq&0,\quad d^2G=\Delta\psi^*_{Ti} {\partial^2G(\psi^0_T)\over \partial\psi^*_{Ti}\partial\psi_{Tj}}\Delta\psi_{Tj} 
\eeqa
For the positivity and negativity of the quadratic forms $d^2R$ and $d^2G$ respectively it suffices to set
\beqa
{\partial^2\over \partial\psi^*_i\partial\psi_j}R(t,\psi^0,u^0)&=&
           \left\{\begin{array}{ll} 
                      0 &\mbox{ for any }  i\neq j\\
                      \delta_i\geq 0 &\mbox{ for all } i=j
                   \end{array}\right. \label{eqKrtv26}\\
{\partial^2\over \partial\psi^*_{Ti}\partial\psi_{Tj}}G(\psi^0_T)&=&
           \left\{\begin{array}{ll} 
                     0& \mbox{ for any } i\neq j\\
                     -\alpha_i\leq 0 &\mbox{ for all } i=j,
                   \end{array}\right.\label{eqKrtv27}
\eeqa  
where $\alpha_i$ and $\delta_i$ are some nonnegative functions.
Inserting the full form of $R$ and $G$ into the above equations yields a set of $n(n+1)/2$  equations of motion for the functions $\sigma_{ij}$, where $n$ is the dimension of the state vector. This means that the dimension of the system grows proportionately to the square of $n$, and for large $n$ the expense of solving  this system normally becomes too high.

In \cite{Konnov99} it is proved that for certain classes of functionals conditions (\ref{eqKrtv12},\ref{eqKrtv13}) can be fulfilled by taking $\phi$ according to eq. (\ref{eqKrtvPhiDef}) with
\beq\label{eqKrtvSgma}
\sigma_{ij}(t)=
           \left\{\begin{array}{ll}
                     \alpha(e^{\gamma(T-t)}-1)+\beta& \mbox{ for }i=j.\\
                     0 &\mbox{ for }i\neq j
                   \end{array}\right.                
\eeq
where $\alpha,\beta<0$ and $\gamma>0$.
Taking $\beta\leq -{\partial^2 F(\psi^0_T)\over \partial\psi^*_T\partial\psi_T}$ always fulfills condition (\ref{eqKrtv27}) and it can be shown that as $\gamma,|\alpha|\to\infty$, for these classes of functionals, condition (\ref{eqKrtv26}) is also fulfilled. The strategy then is to begin with $\sigma=0$ and if the objective does not  decrease, take increasingly larger $\gamma, |\alpha|, |\beta|$ until a decrease in the objective is achieved.
\section{\label{sec:ApplNLSE}Application to the NLSE}
\subsection{General formulation}
We wish to apply this algorithm to optimizing a quantum system governed by the nonlinear Schr\"odinger equation
\beq \label{eqKrtvNLSE}
\ket{\dot\psi}=-i\hat H_{NL}(|\psi|^2)\ket{\psi}=-i(\hat H+\mu|\psi|^2)\ket{\psi}, 
\eeq
where $\hat H=\hat K+\hat V$ is the usual linear Hamiltonian operator consisting of kinetic and potential parts and $\mu$ is the coefficient of the additional non linear term.  The objective is defined as for the linear case, as minimizing
\beq
J=-\braket{\psi_T}{\hat P|\psi_T}+\lambda\int dt f^0(u).
\eeq
In realization of step 1 of the iterative method we choose $\phi(t,\psi)=\braket{\chi}{\psi}+\braket{\psi}{\chi}+{1\over 2}\braket{\Delta\psi}{\sigma|\Delta\psi}$ and find the Hamiltonian of the system to be 
\beqa
H(t,\psi,u,\chi)&=&-i\braket{\chi}{\hat H_{NL}|\psi}+i\braket{\psi}{\hat H^\dagger_{NL}|\chi}-\lambda f^0(u)\nonumber\\
&=&2\Im\braket{\chi}{\hat H_{NL}|\psi}-\lambda f^0(u),
\eeqa
just as in the linear case. However it must be remembered that here $\hat H_{NL}$ depends on $\psi$ and $\psi^*$, and therefore using eq. (\ref{eqKrtv19}) yields for the conjugate vector
\beqa
\ket{\dot\chi}&=&-{\partial\over\partial\psi^*}H(t,\psi^0,u^0,\chi)\nonumber\\
              &=&i\mu(\psi^0)^2\ket{\chi^*}-i(\hat H^\dagger+2\mu|\psi^0|^2)\ket{\chi}.\label{eqKrtvChiNL}
\eeqa
Note that this equation differs in two respects from its linear counterpart (eq. \ref{eqKrtvChiL}). First, it involves $\ket{\psi^0}$, which does not cause special problems except that the vector  $\ket{\psi^0}$ must be stored and used in propagating $\ket{\chi}$. The more cumbersome difficulty arises from the fact that the equation obtained is no longer linear in  $\ket{\chi}$, but rather contains an extra term linear in  $\ket{\chi^*} (=\bra{\chi})$. Nevertheless the coupled equations
\beq
{\partial\over\partial t}\left(\begin{array}{c}\ket{\chi}\\ \ket{\chi^*}\end{array}\right)
=-i\left\{\left(\begin{array}{cc}\hat K & 0\\ 0 &-\hat K \end{array} \right)
         +\left(\begin{array}{cc}a & -b\\ b^* & -a \end{array}\right)\right\}
\left(\begin{array}{c}\ket{\chi}\\ \ket{\chi^*}\end{array}\right),
\eeq
where $\hat K=-{\hbar^2\over 2m}{\partial^2\over \partial x^2}$, $a=\hat V(x)+2\mu|\psi^0|^2$ and $b=\mu(\psi^0)^2$, or in matrix form
\beq
\dot{\vec\chi}=-i(K+V)\vec\chi,
\eeq
can be solved using the split operator method for separating the treatment of the kinetic and potential parts of the Hamiltonian: $\vec\chi(t+\delta t)=e^{-iK\delta t/2}e^{-iV\delta t}e^{-iK\delta t/2}\vec\chi(t)$. The kinetic evolution can be computed using the usual Fourier transform methods and the potential part can be evaluated by diagonalization to give,
\begin{widetext}
\beq
e^{-iV\delta t}={1\over D}\left(\begin{array}{cc}
{|b|^2\over a-D}\cos(D\delta t)-ae^{iD\delta t}& ib\sin(D\delta t)\\
 -ib^*\sin(D\delta t)& {|b|^2\over a-D}\cos(D\delta t)-ae^{-iD\delta t}
\end{array}\right),
\eeq
with $D=\sqrt{a^2-|b|^2}$.
Finally, introducing the complex conjugation operator $\hat C$ and the Fourier transform operator $\hat Z$, this procedure yields the following formula for the numerical propagation step
\beq
\ket{\chi(t+\delta t)}=\hat Z^\dagger e^{i{\hbar^2 k^2\over 4m}\delta t}\hat Z\left\{{1\over D}\left({|b|^2\over a-D}\cos(D\delta t)-a\exp(iD\delta t)\right)+i{b\over D}\sin(D\delta t)\hat C\right\}\hat Z^\dagger e^{i{\hbar^2 k^2\over 4m}\delta t}\hat Z\ket{\chi(t)}.
\eeq
\end{widetext}

Having obtained $\ket{\chi(t)}$ for all $t$, we now proceed to realize step 2 of the algorithm and according to eq. (\ref{eqKrtv11}), find for each point in time a field which maximizes $H(t,\psi,u,{\partial\phi\over\partial\psi})=2\Im\braket{\chi+{1\over 2}\sigma\Delta\psi}{\hat H(u)|\psi}-\lambda f^0(u)$. Or mathematically
\beq
\tilde u(t,\psi)=arg \max_u H(t,\psi,u,{\partial\phi \over \partial \psi}).
\eeq

Step 3 of the algorithm is fulfilled simultaneously with step 2, by  simultaneously propagating $u$ and $\psi$ such that each new $u(t)$ is used directly in propagating $\psi(t)\to\psi(t+\delta t)$.  For $\sigma$ we use formula (\ref{eqKrtvSgma}) and according to the algorithm described in the previous section we begin with $\sigma=0$ and increase the parameters $\gamma,|\alpha|,|\beta|$ until achieving decrease of the objective.
\subsection{Application to a concrete problem}
\begin{figure}[hbt]
\centerline{
   \mbox{\epsfig{figure=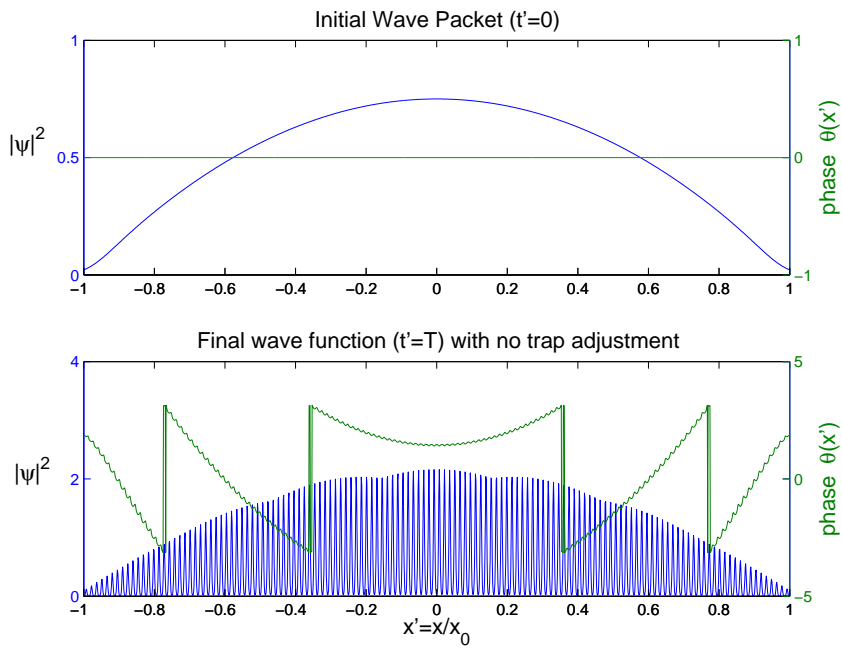,width=4.4in}}
   }
   \caption[IniWp]{\label{IniWp}Wave packet at t=0,and t=T with no trap adjustments.}
\end{figure}

We consider a 1D Bose Einstein Condensate (BEC) confined by a harmonic trap and governed by the Gross-Pitaevskii equation
\beq
\ket{\dot\psi}=-i(\hat K+\hat V+NU_0|\psi|^2)\ket{\psi},
\eeq 
where $\hat K$, $\hat V$ are as above and $NU_0$ is the nonlinear atom-atom interaction strength, $N$ being the number of atoms.  
The BEC is initially in the ground state of the trap potential and is therefore stationary. An optical lattice is then switched on, having the effect of separating the BEC wave packet into a series of localized pieces. The potential energy operator therefore takes the form $\hat V=Kx^2+S(t)V_0\cos^2(kx)$, where $K$ is the trap constant, $k$ is the laser field wave number, $V_0$ is the lattice intensity and  the switching on function of the field is denoted $S(t)$. In applications to quantum computing, these localized wave packets are to represent quantum bits. However, due to the nonlinearity of the equations, the condensate develops a phase that varies from lattice site to lattice site (see figure \ref{IniWp}), which is undesirable for quantum computing, since these algorithms assume that there is zero relative phase among the  various single quantum bits.
The problem is therefore to eliminate this phase profile by adjusting the trap strength during the switching on of the laser field. From the OCT perspective the trap constant $K(t)$ is taken as the control and the objective is to minimize the variance of the phase of the wave packet, $\theta(x)$, at some final time $T$. The phase being a multivalued function poses problems; we therefore consider instead minimizing the variance of $\cos(\theta)={1\over 2}{\psi+\psi^*\over|\psi|}$ such that the objective becomes
\beqa
J&=&\average{\cos^2(\theta_T)}-\average{\cos(\theta_T)}^2\nonumber\\
&=&\braket{\psi}{\cos^2(\theta_T)|\psi}-\braket{\psi}{\cos(\theta_T)|\psi}^2
\eeqa 
Using the first part of eq. (\ref{eqKrtv19}) we get the equations of motion for $\chi$ as in eq. (\ref{eqKrtvChiNL}) and taking a derivative of $J$ with respect to $\psi_T^*$ we get according to (\ref{eqKrtv19}), the boundary conditions:
\beq
\chi_T=-{\partial J\over \partial \psi_T^*}=-\Re[\psi_T]+{1\over 2}|\psi_T|\average{\cos(\theta_T)}({\psi_T\over\psi_T^*}+3).
\eeq
The Hamiltonian of the problem is $H(t,\psi,K,\chi)=K(t)\braket{\chi+{1\over 2}\sigma(\Delta\psi)}{x^2|\psi}$ so that according to the above procedure we improve $K$ at each iteration by $K\to K+\lambda^{-1}\braket{\chi+{1\over 2}\sigma(\Delta\psi)}{x^2|\psi}$.
  
\subsection{Optimization Results} 

Following \cite{Trippenbach00} we transform the NLSE to dimensionless units $t'=t/t_0$ and $x'=x/x_0$ where $x_0=x_{TF}=20.3\mu m$ and $t_0={mx_0^2\over2\hbar}=75ms$. 
The Thomas-Fermi radius $x_{TF}=\sqrt{{2\mu_{TF}/m\omega^2_{trap}}}$ gives the size of the condensate and is defined through a chemical potential $\mu_{TF}\equiv\hbar/t_{NL}$ determined by normalization of the wave function to unity. The wave function too is scaled $\psi\to\sqrt{x_{TF}}\psi$ and in order to keep the time scales of our 1D model comparable to the 3D reality we adjust $U\to CU$ by a factor $C={\sqrt{\pi}\over\Gamma(2+1/2)}={4\over 3}$ \cite{Trippenbach00}. We take $t_{NL}=96.2\mu s$, optical wavelength $\lambda=589nm$ and $V_0=10.94E_R$ for the final field intensity. All parameters were taken to resemble the experiments described in  \cite{Trippenbach00}. Performing these transformations we end up finally with a dimensionless NLSE,
\beq
i\ket{\dot\psi}=\left(-{1\over 4}{\partial^2\over\partial x^2}+K(t)x^2+S(t)V\cos^2(kx)+U|\psi|^2\right)\ket{\psi},
\eeq 
where the trap constant  $K=\omega_{trap}^2t_0^2$, the field Intensity $V=V_0{t_0/\hbar}=2\times 250^2$ and the nonlinear coefficient  $U={4\over 3}\mu_{TF}{t_0/\hbar}=1039$, such that all space, time and energy quantities are now expressed in units of $x_0$, $t_0$ and $\hbar/t_0$ respectively.

\begin{figure}[hbt]
       {\epsfig{figure=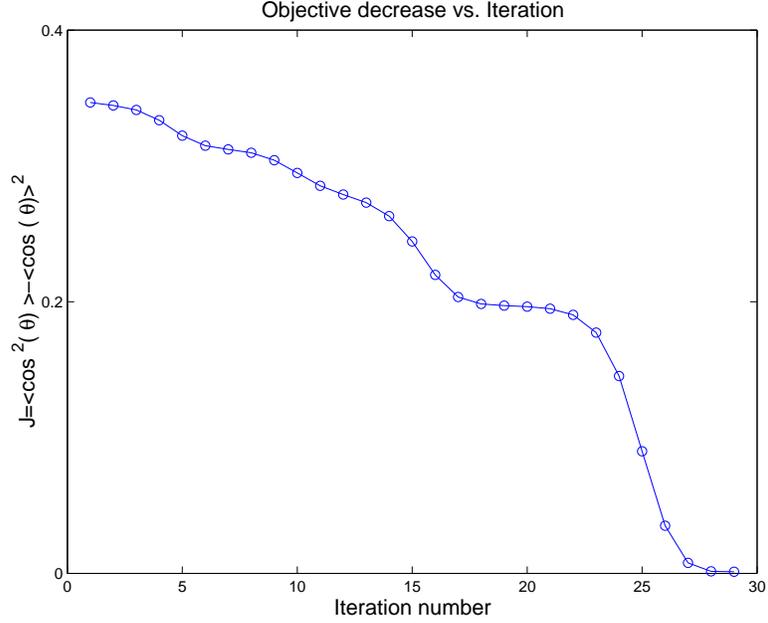,width=4in}}  
   \caption[OptObj]{\label{OptObj}Objective decrease as function of iteration.}
\end{figure}

\begin{figure}[hbt]
   \epsfig{figure=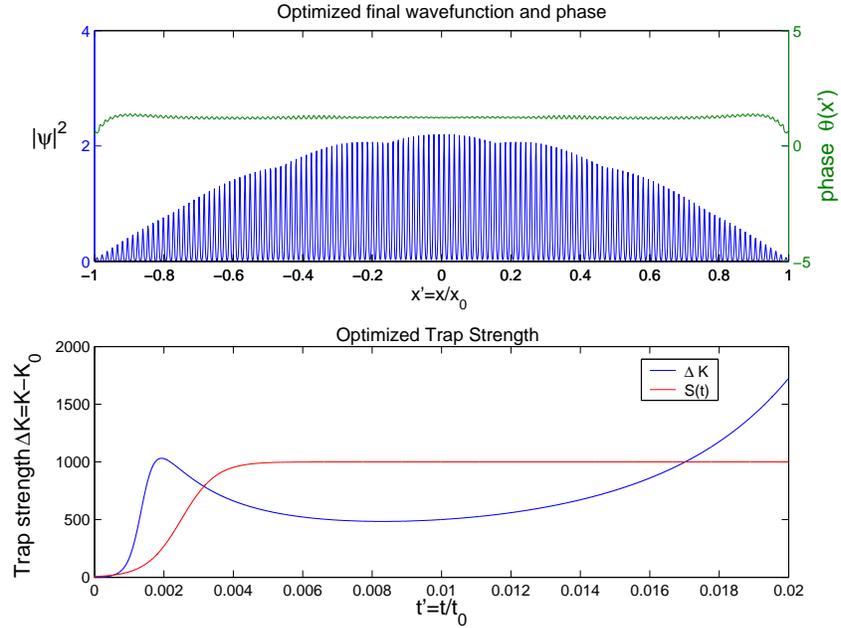,width=4.4in}
   \caption[OptWp]{\label{OptWp}Optimized trap strength evolution (bottom), and final wave packet at t=T (top). The flat phase is strikingly apparent.}
\end{figure}

Initially the wave packet is in an eigenstate of the potential with trap constant $K_0=779$ and is therefore static. The switching on function plotted in figure \ref{OptWp}, turns on the optical potential at a quarter of the optimizing interval ($T/4$) and is constant from there to the final time. 
With no adaption of the trap constant a phase develops across the wave function as shown in figure \ref{IniWp}. The optimization process decreases the objective monotonically as plotted in figure \ref{OptObj} and yields a strategy of increasing $\Delta K(t)=K(t)-K_0$ to achieve a flat phase at the final time, $T=1500\mu s$. These striking results are shown in figure \ref{OptWp}.

\section{\label{sec:Conclude}Conclusions} 
Using a set of general methods developed by Krotov we have extended the capabilities of Optimal Control Theory 
to the Nonlinear Schr\"odinger Equation (NLSE).  Although the linearized version of the method
is sufficient for the linear Schr\"odinger equation, the full flexibility of the general method is required for 
a rigorous treatment of the nonlinear Schr\"odinger equation.  
Mention should be made of the interesting recent work of Hornung and de Vivie-Riedle
\cite{Hornung}, applying optimization techniques to molecule formations in a BEC, although the $\phi$ function in that work included linear terms only.  
A parallel approach was pursued by P\"otting {\it et al} \cite{Potting01} who used  genetic algorithms to control the momentum state of a BEC.
The significance of this paper is twofold.  First, both formal equations and a concrete and efficient algorithm were 
presented for optimizing the NLSE in cases where the nonlinear terms are significant.  
Second, the methodology was applied successfully to an interesting physical problem
confronting the use of trapped Bose-Einstein condensates (BECs) for quantum computing, namely producing
a constant final phase profile across the condensate after an optical lattice is turned on. 
Further work on understanding analytically the mechanism found by OCT is still underway. 
We believe the working equations developed here will have many more applications in systems governed by the NLSE, 
including both BECs and soliton fiber optics. 
\begin{acknowledgments}
We wish to thank Vadim F. Krotov and Vladimir A. Kazakov for their useful correspondence advice  and references. We thank Carl Williams for his hospitality at NIST Gaithersburg where some of this work begun. We also thank Thomas Hornung for his comments and discussions.
This work was supported by 
the US Office of Naval Research (grant No. N00014-01-1-0667)
the Israel Science Foundation (grant No. 128/00-2),
and the German-israel BMBF (grant No. 13N 7947).
\end{acknowledgments}   

\bibliographystyle{phaip}
\bibliography{shlomo}

\end{document}